\documentclass[11pt]{article}
\usepackage[top=1.in,bottom=1.in,left=1.in,right=1.in]{geometry}
\usepackage{graphicx}
\usepackage{wrapfig} 
\usepackage{subfig} 
\usepackage{amsmath}
\usepackage{color}
\usepackage[pagewise]{lineno}
\graphicspath{{figures/}}
\DeclareGraphicsExtensions{.pdf,.png,.jpg}

\newcommand{\nub}{\bar{\nu}}
\newcommand{\nue}{\nu_e}
\newcommand{\nueb}{\bar{\nu}_e}
\newcommand{\numu}{\nu_\mu}
\newcommand{\numub}{\bar{\nu}_\mu}

\begin{document}
\title{A new investigation of electron neutrino appearance oscillations 
with improved sensitivity in the MiniBooNE+ experiment}
\author{
\large{\textbf{MiniBooNE+ Collaboration}} 
\vspace*{5pt} \\
R. Dharmapalan, S. Habib, C. Jiang, \& I. Stancu \\
\textit{University of Alabama, Tuscaloosa, AL 35487} \vspace{3pt} \\
Z. Djurcic \\
\textit{Argonne National Laboratory, Argonne, IL 60439} \vspace{3pt} \\
R. A. Johnson \& A. Wickremasinghe \\
\textit{University of Cincinnati, Cincinnati, OH 45221} \vspace{3pt} \\
G. Karagiorgi \& M. H. Shaevitz \\
\textit{Columbia University, New York, NY 10027} \vspace{3pt} \\
B. C. Brown, F.G. Garcia, R. Ford, W. Marsh, C. D. Moore, \\
D. Perevalov,  \& C. C. Polly \\
\textit{Fermi National Accelerator Laboratory, Batavia, IL 60510} \vspace{3pt} \\
J. Grange,  J. Mousseau, B. Osmanov, \& H. Ray \\
\textit{University of Florida, Gainesville, FL 32611} \vspace{3pt}\\
R. Cooper, R. Tayloe, R. Thornton \\
\textit{Indiana University, Bloomington, IN 47405}\vspace{3pt}  \\
G. T. Garvey, W. Huelsnitz, W. C. Louis, C. Mauger, G. B. Mills, \\
Z. Pavlovic, R. Van de Water, \& D. H. White \\
\textit{Los Alamos National Laboratory, Los Alamos, NM 87545} \vspace{3pt} \\
R. Imlay, M. Tzanov \\
\textit{Louisiana State University, Baton Rouge, LA 70803}\vspace{3pt} \\
B. P. Roe \\
\textit{University of Michigan, Ann Arbor, MI 48109} \vspace{3pt} \\
A. A. Aguilar-Arevalo \\
\textit{Instituto de Ciencias Nucleares, Universidad Nacional Aut\'onoma de M\'exico,} \\
\textit{M\'exico D.F. M\'exico} \vspace{3pt} \\
T. Katori \\ 
\textit{Queen Mary University of London, London, E1 4NS, United Kingdom} \vspace{3pt} \\
P. Nienaber \\
\textit{Saint Mary's University of Minnesota, Winona, MN 55987}\\
}
\maketitle


\begin{abstract}
We propose the addition of scintillator to the existing MiniBooNE detector to allow a test of the neutral-current/charged-current (NC/CC) nature of the MiniBooNE low-energy excess.  
Scintillator will enable the reconstruction of 2.2~MeV $\gamma$s from neutron-capture on protons following neutrino interactions.   
Low-energy CC interactions where the oscillation excess is observed should have associated neutrons with less than a 10\% probability.  
This is in contrast to the NC backgrounds that should have associated neutrons in approximately 50\% of events.  
We will measure these neutron fractions with $\nu_\mu$ CC and NC events to eliminate that systematic uncertainty.
This neutron-fraction measurement requires $6.5\times10^{20}$ protons on target delivered to MiniBooNE with scintillator added in order to increase the significance of an oscillation excess to over $5\sigma$.    

This new phase of MiniBooNE will also enable additional important studies such as the spin structure of nucleon ($\Delta s$) via NC elastic
scattering, a low-energy measurement of the neutrino flux via $\numu~^{12}C \rightarrow \mu^{-}~^{12}N_\textrm{g.s.}$ scattering, and a test of the quasielastic assumption in neutrino energy reconstruction.  
These topics will yield important, highly-cited results over the next 5 years for a modest cost, and will help to train Ph.D. students and postdocs.  
This enterprise offers complementary information to that from the upcoming liquid Argon based MicroBooNE experiment.  
In addition, MicroBooNE is scheduled to receive neutrinos in early 2014, and there is minimal additional cost to also deliver beam to MiniBooNE.
\end{abstract}

\section{Introduction}
The MiniBooNE experiment has, for the last 10 years, searched
for $\numu\rightarrow\nue$ and $\numub\rightarrow\nueb$ 
in the Booster Neutrino Beamline at Fermilab.  The beam was shut down in April 2012
to enable the Fermilab accelerator complex to be upgraded in preparation for 
delivering higher beam power to the NOvA experiment.  Before the shutdown, 
MiniBooNE completed an antineutrino phase of running which brought the total 
amount of beam delivered to 
the experiment to $11.3\times 10^{20}$ protons on target (POT) in antineutrino
mode and $6.5\times 10^{20}$~POT in neutrino mode.  

\begin{figure}[h]
\centering
\begin{minipage}[t]{0.48\textwidth}
\centering
{\includegraphics[width=0.95\textwidth]{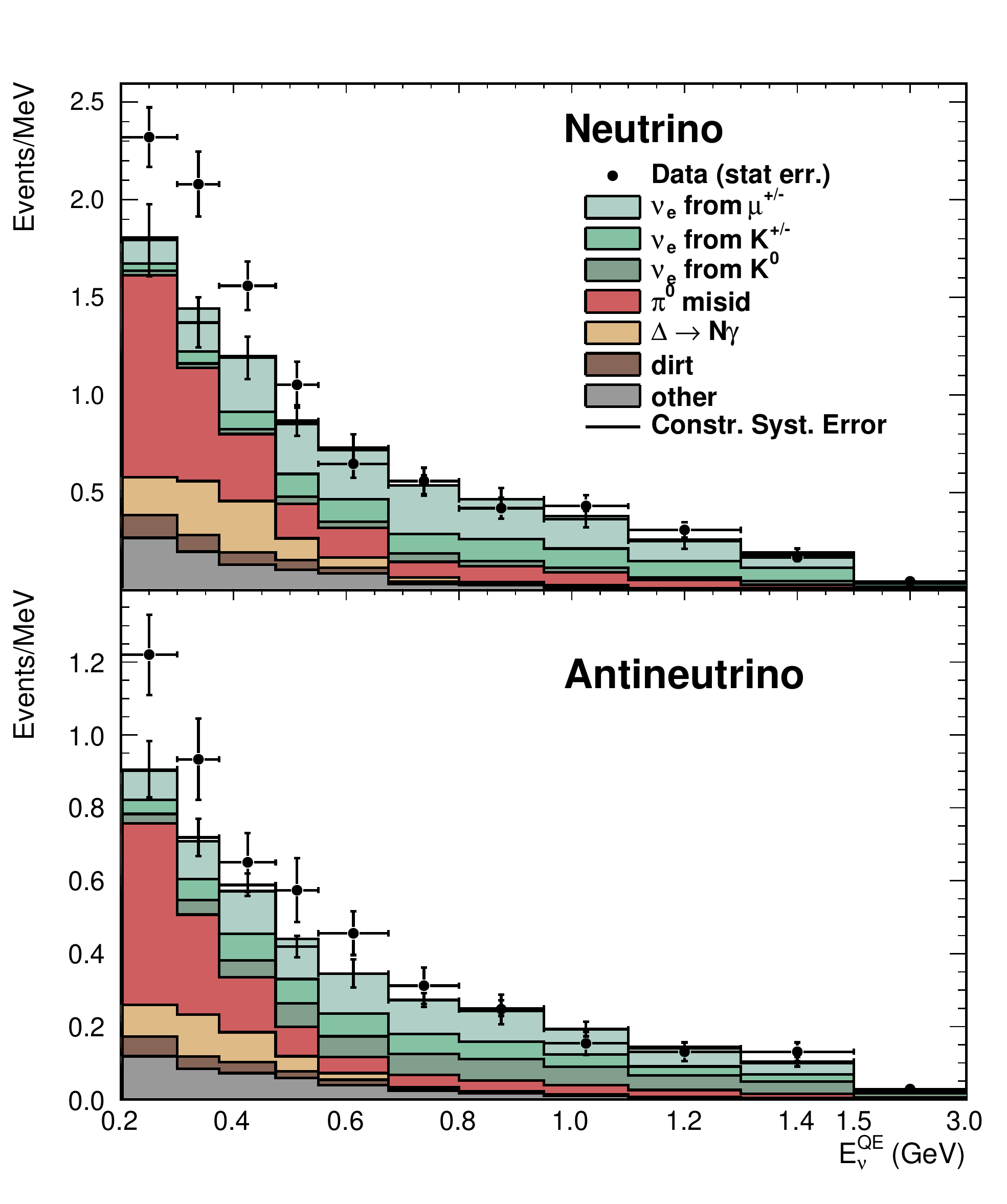}}
\caption{The neutrino mode (top) and antineutrino mode (bottom) 
reconstructed neutrino energy, $E_\nu^{QE}$, distributions   
for data (points with statistical errors) 
and predicted background (histogram with systematic errors).}
\label{fig:nu_nubar_stack}
\end{minipage}
\hspace{3mm}
\begin{minipage}[t]{0.47\textwidth}
\centering
{\includegraphics[width=0.95\textwidth]{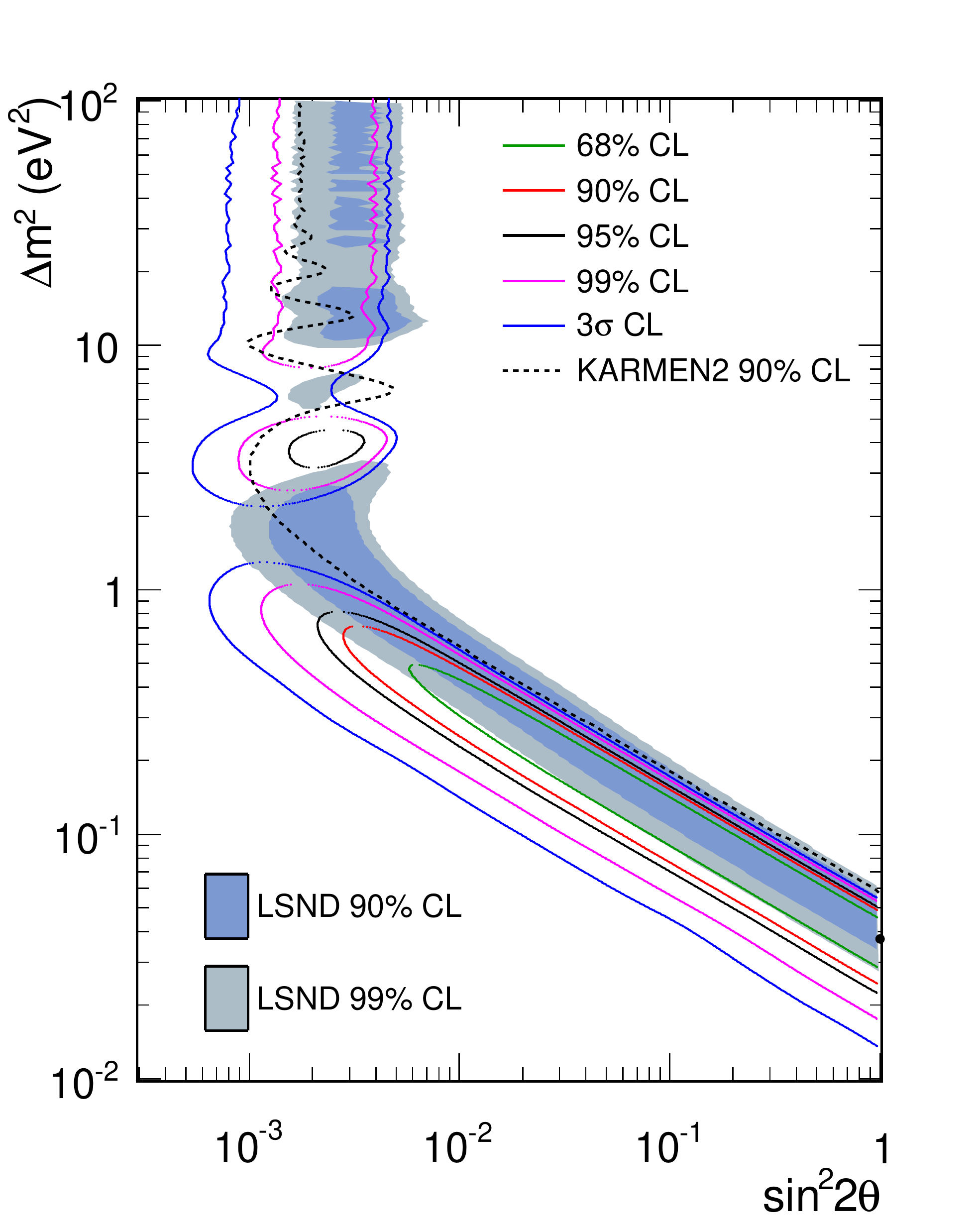}}
\caption{
MiniBooNE allowed regions in combined neutrino and antineutrino mode for events with
$200<E^{QE}_{\nu}< 3000$~MeV within a 2$\nu$ oscillation model. }
\label{fig:mb_nu_nubar_combined_fit}
\end{minipage}
\end{figure}

Both $\numu\rightarrow\nue$ and $\numub\rightarrow\nueb$ oscillation 
analyses have been conducted with this data 
individually~\cite{AguilarArevalo:2007it}-\cite{AguilarArevalo:2010wv}
and recently as a combined data set with the latest updates to the
antineutrino data~\cite{AguilarArevalo:2012va}.
There is an excess of events over the calculated background in both modes 
(Fig.~\ref{fig:nu_nubar_stack}) examined individually as well as for the combined 
data set which contains a total excess of  $240.3\pm62.9$ ($3.8\sigma$) events.  
A two-neutrino fit to the combined data set  yields allowed parameter 
regions (Fig.~\ref{fig:mb_nu_nubar_combined_fit}) which
are consistent with oscillations in the 0.01 to 1 eV$^2$ $\Delta m^2$ range 
and are consistent with the regions reported by the LSND experiment~\cite{Aguilar:2001ty}.

The excess occurs in both $\nu$ and $\nub$ samples at low energy, and so it is 
natural to consider more carefully the largest backgrounds in 
that energy region.  They are dominated by neutral-current (NC) $\pi^0$ and NC
$\Delta$ radiative decays ($\Delta \rightarrow N \gamma$). Both of these NC 
processes are constrained  by measurements within MiniBooNE, but an anomalous 
process such as NC $\gamma$  production, not sufficiently accounted for in the MiniBooNE 
analysis, could lessen the significance of the oscillation excess.  
We are proposing to measure these backgrounds with a new technique combined with
additional running of MiniBooNE. 

The MiniBooNE detector uses 800 tons of mineral oil
(CH$_2$) as a target medium for inducing CC and NC neutrino interactions.  
The mineral oil also serves as the detector medium for observing the final state 
particles resulting from the interactions.
This is achieved via detection of the Cerenkov light from charged particles in the 1280 
8'' photomultiplier tubes (PMTs) that line the inside of the spherical
detector tank.   In addition to the Cerenkov light produced in a cone
around the trajectory of charged particles, some isotropic 
scintillation light is produced due to the presence of aromatic impurities
in the mineral oil.  

We propose adding approximately 300~kg of PPO scintillator to the 800 tons 
of MiniBooNE mineral oil to increase the amount of scintillation light produced
by 2.2~MeV $\gamma$ that result from delayed ($\tau\approx186~\mu$s) neutron 
capture on protons within the mineral oil.  This will allow an important test of the 
oscillation signal by checking that the excess is indeed due to CC interactions
of low-energy neutrinos and not an incorrectly calculated NC background.  This 
can be done by counting  $n$-capture events that follow oscillation candidate
events.  If the excess is indeed due to CC interactions of low energy $\nue$, 
only approximately $10\%$ of the excess will have associated $n$-capture events.  
If, instead, the excess is due to a NC process, one would expect many
more neutrons produced since the interactions are caused from
higher energy neutrinos.  One expects approximately 50\% of NC
background events to have an associated neutron.  An attractive
feature of this method is that the neutron fraction for CC and NC
processes may be measured with MiniBooNE via similar channels, thereby
eliminating that systematic uncertainty.

The increased level of scintillation will enable several other important
measurements.  The detection of $n$-capture enables a measurement of
the neutron to proton ratio in NC elastic scattering which is sensitive to
the strange-quark spin of the nucleon ($\Delta s$).  The  $\beta$ decay from 
the $^{12}N_\textrm{g.s.}$ in the $\numu~^{12}C \rightarrow \mu^{-}~^{12}N_\textrm{g.s.}$ process
will be better reconstructed which will allow a measurement of this process
and a check of the low-energy neutrino flux.   Low-energy recoil nucleons
will be more visible within neutrino events allowing a test of the quasielastic
assumption in neutrino energy reconstruction.


\section{Physics Goals}
A main motivation for adding scintillator to MiniBooNE is to provide a test of 
the nature of the low-energy excess of events observed in both the $\nue$ and $\nueb$ appearance 
searches conducted by MiniBooNE.  The addition of scintillator will also enable 
an investigation of the strange-quark contribution  to the nucleon spin ($\Delta s$), 
a measurement of the $\numu~^{12}C \rightarrow \mu^{-}~^{12}N$ reaction, and a test 
of the quasielastic assumption in neutrino energy reconstruction. 

\subsection{Oscillation search with CC/NC identification}

MiniBooNE has measured a $3.8\sigma$ excess of oscillation candidate events in the combined
$\numu$ and $\numub$ data sets collected to date at Fermilab~\cite{AguilarArevalo:2012va}.
As can be seen in Fig.~\ref{fig:nu_nubar_stack}, the predicted backgrounds in the 
low energy regions, where the excess is most substantial, are dominated by neutral current
backgrounds.  These NC backgrounds are from two major sources:  misidentification 
of the $\pi^0$ (``$\pi^0$ misid'') and the production of $\Delta$ baryons which then radiatively
decay (``$\Delta \rightarrow N\gamma$'').   A test of these NC backgrounds in a measurement 
with different systematic errors would be quite valuable to firmly establish the oscillation excess.

\begin{figure}[h]
\centering
\subfloat[$\nue CCQE$]{
\label{fig:fd_CCNC_set:a}
{\includegraphics[width=0.31\textwidth]{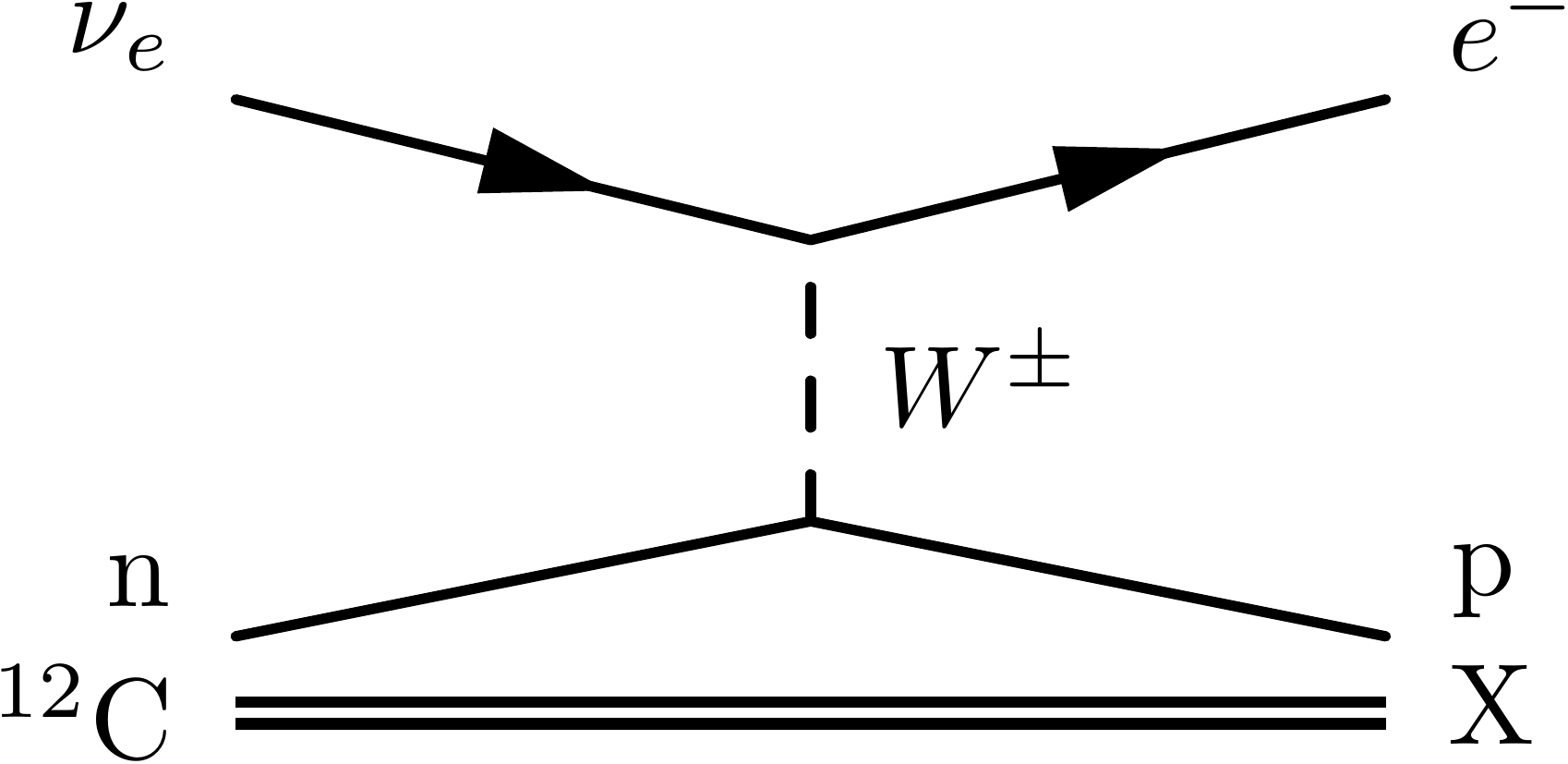}}
}
\hspace{-5mm}
\subfloat[$\numu NC \pi^0$ ]{
\label{fig:fd_CCNC_set:b}
{\includegraphics[width=0.31\textwidth]{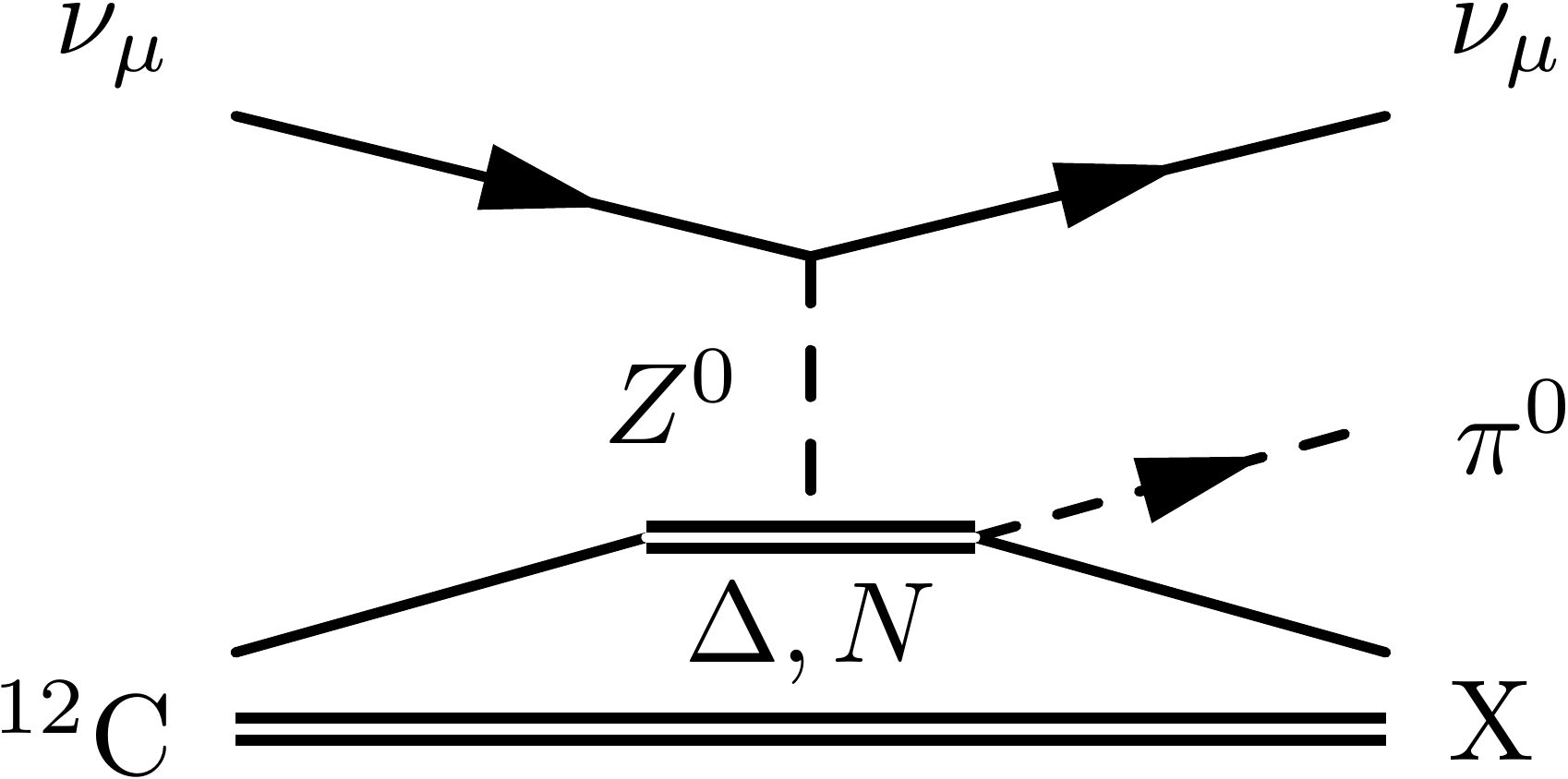}}
}
\hspace{-5mm}
\subfloat[$\numu NC \gamma$]{
\label{fig:fd_CCNC_set:c}
{\includegraphics[width=0.31\textwidth]{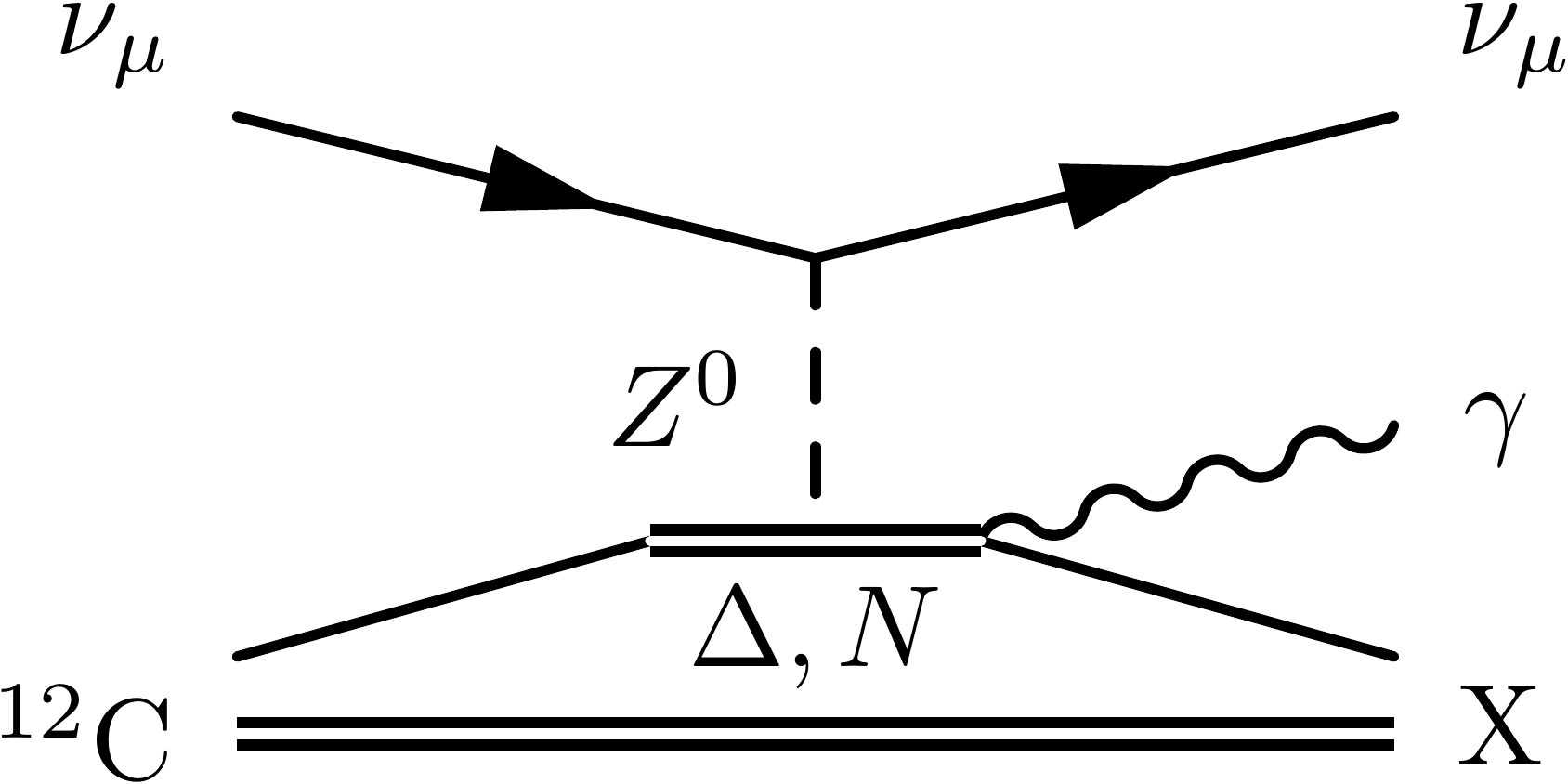}}
}
\caption{Diagrams of signal (a) and background (b,c) neutrino oscillation candidate events.}
\label{fig:fd_CCNC_set}
\end{figure}

MiniBooNE can perform this test by detecting neutrons associated with oscillation candidate
events. In brief, at low $E_\nu^{QE}$, true CC oscillation events (Fig.~\ref{fig:fd_CCNC_set:a}) 
should contain final-state neutrons in less than 10\% of the events while the 
NC backgrounds (Figs.~\ref{fig:fd_CCNC_set:b},\ref{fig:fd_CCNC_set:c}) 
should contain neutrons  in $\approx 50\%$ of the events.

Note that  $E_\nu^{QE}$ is the reconstructed neutrino energy using the assumption
of neutrino quasielastic scattering from a neutron. This quantity should be a good 
estimate for the true neutrino energy in true CC oscillation events (excepting possible
nuclear effects, Sec.~\ref{sec:enuqetest}).
However, because of the large missing energy in NC events, the true neutrino energy is,
on average, much higher than $E_\nu^{QE}$ for NC backgrounds events.  So, in a first approximation, the NC 
backgrounds will contain more final state neutrons because the events are from higher true neutrino
energy. More energy is transferred to the nucleus which causes more neutron
production as compared to the CC signal in which the true neutrino energy is lower and 
(neglecting final state interactions) produces a single proton. 

In practice, we would rerun the MiniBooNE oscillation search in neutrino mode after
the addition of scintillator in order to enable neutron detection. Oscillation candidates
would be selected with the same strategy as the original search.  From this sample, we would 
search for neutron capture events and measure the neutron fraction which would test the
NC background estimates. An important feature of this measurement is that the neutron fraction
may be ``calibrated'' for the oscillation search via separate MiniBooNE $\numu$ CCQE and 
$\numu$ NC$\pi^0$ measurements which
greatly reduces errors from any nuclear model uncertainties.

\subsection{Proton to neutron ratio in NC elastic events}
The NC neutrino-nucleon elastic scattering (NC elastic) interaction, $\nu N \rightarrow \nu N$, 
is sensitive to the isoscalar-axial structure of the nucleon~\cite{Garvey:1993sg}, so  
will be sensitive to the effects of strange-quark contributions to the nucleon spin
($\Delta s$).  Therefore, the right type of measurement of NC elastic scattering
would contribute substantially to the nucleon spin puzzle, an area of continued interest and
effort (e.g.~\cite{Bass:2009dr}). This measurement of NC elastic scattering has not yet been realized. 

MiniBooNE has made the most accurate measurement to date of the differential cross section for
$\nu N \rightarrow \nu N$~\cite{AguilarArevalo:2010cx} and the analysis for the 
$\bar{\nu} N \rightarrow \bar{\nu} N$ process is almost complete~\cite{Dharmapalan:2011sa}. 
While these are valuable measurements to help with understanding of neutrino-nucleon scattering, 
they are not
sensitive to $\Delta s$ because MiniBooNE is not able currently to distinguish between 
neutrons and protons.   The $\nu p \rightarrow \nu p$ process is sensitive to $\Delta s$
with the opposite sign as $\nu n \rightarrow \nu n$ and any strange quark effects 
cancel in the existing MiniBooNE measurement.

This situation changes abruptly with the addition of neutron-capture tagging.  In that case,
the neutrons and protons can be separately identified and the neutron/proton
ratio,
\begin{equation}
R(p/n) = \frac{\sigma(\nu p \rightarrow \nu p)}{\sigma(\nu n \rightarrow \nu n)},
\label{eq:rpn}
\end{equation} 
is quite sensitive to $\Delta s$~\cite{Garvey:1993sg}.  Based on previous studies~\cite{Bugel:2004yk},
a rough estimate is that a 10\% measurement of $R(p/n)$ should result in an error of $\approx 0.05$ uncertainty
on $\Delta s$.  It should be realized that the recent results from MiniBooNE on the unexpectedly large 
CCQE cross section~\cite{AguilarArevalo:2010zc} may call into question the theoretical uncertainty involved 
in extracting $\Delta s$ from $R(p/n)$. If there are multinucleon correlations contributing substantially
to NC elastic scattering, it may not be clear how that affects the extraction of $\Delta s$. 
Regardless, a 10\% 
measurement of $R(p/n)$ will be a valuable constraint and will further more theoretical investigation.

\subsection{A measurement of $\numu~^{12}C \rightarrow \mu^{-}~^{12}N_\textrm{g.s.}$}
The reaction $\numu~^{12}C \rightarrow \mu^{-}~^{12}N_\textrm{g.s.}$ is an interesting reaction
to study with a scintillator-enhanced MiniBooNE for several reasons.   
It comes with a  distinctive tag from the $\beta$-decay of the $^{12}N_\textrm{g.s.}$ with endpoint 
energy of 16.3~MeV and lifetime of 15.9~ms.  This addition of scintillator to MiniBooNE will allow
for high efficiency and better reconstruction of the $\beta$-decay.
Since it is an exclusive reaction, the theoretical cross section can be calculated to 
$\approx 2\%$ very near threshold~\cite{Athanassopoulos:1997rn}. It was measured by 
LSND for both $\numu$ and $\nue$~\cite{Athanassopoulos:1997rn,Athanassopoulos:1997rm} 
and by KARMEN for $\nue$~\cite{Bodmann:1992ur} to agree with theory to within experimental
errors.  A measurement by MiniBooNE of this theoretically 
well-known reaction would enable a test of the low-energy neutrino flux which could better constrain 
the low-energy oscillation excess.

The event signature is quite distinct.  The low-energy prompt $\mu^{-}$ and subsequent
decay $e^{-}$ would be detected with the usual techniques employed for $\numu$ CCQE events
combined with a requirement for a detected $\beta$-decay candidate.  With the 
addition of scintillator to make 2.2~MeV $\gamma$ visible, the efficiency for detecting
the 16.3~MeV-endpoint $\beta$ will be quite high.

The challenge is that the fraction of the total $\numu$ scattering events that interact
via $\numu~^{12}C \rightarrow \mu^{-}~^{12}N_\textrm{g.s.}$ is small.  In the lowest energy
bin at $E_\nu \approx 250$~MeV, the cross section is about 4\% of the $\numu$ CCQE cross section,
falling to about  0.5\% by $E_\nu \approx 400$~MeV~\cite{Kolbe:1999au}.  However, with the 
data sample proposed here the total $\numu$ event sample will be large, the $^{12}N_\textrm{g.s.}$ 
signature quite distinct, and an analysis will be worth the effort.

\subsection{A test of the QE assumption in neutrino energy reconstruction}
\label{sec:enuqetest}
MiniBooNE has reported absolutely normalized cross sections for various $\numu$-carbon processes
including  $\numu$ CCQE~\cite{AguilarArevalo:2010zc}, CC$\pi^+$~\cite{AguilarArevalo:2010bm}, 
CC$\pi^0$~\cite{AguilarArevalo:2010xt}, NC elastic~\cite{AguilarArevalo:2010cx}, 
and NC$\pi^0$~\cite{AguilarArevalo:2009ww}.  
They all show a 30-40\% larger cross section than predicted in previously existing 
models (e.g.~\cite{Caballero:2005sj}).   
One emerging idea is that two-nucleon correlations in carbon are contributing significantly to the
interaction cross section~\cite{Martini:2009uj,Martini:2010ex}.  If this is the correct explanation
for the extra strength in these neutrino interactions, then it could also have a significant
effect on the reconstructed neutrino energy in oscillation events, $E_\nu^{QE}$, which assumes quasielastic 
scattering from single nucleons within carbon~\cite{Martini:2012fa}.  In short, the reconstructed 
neutrino energy may be incorrect in a large fraction of the oscillation events leading  to incorrect conclusions
about the resulting fits to oscillation models.

The addition of scintillator will allow this idea to be experimentally tested.  With the 
scintillator addition proposed here,
the detector response to final state nucleons in a typical CCQE event will be increased by
about a factor of five.  This scintillation light is a measure of the total energy in the 
event ($E_\nu^{total}$) as opposed to that reconstructed from the just the lepton track, $E_\nu^{QE}$.
A comparison of $E_\nu^{QE}$ with $E_\nu^{total}$ will allow further insight into the two-nucleon
correlation issue in general and, specifically, into its relevance to the low-energy oscillation excess. 

\section{Implementation}
In this section, we explore some details of how to prepare and run MiniBooNE with the addition
of scintillator.  

\subsection{Suggested plan for adding scintillator}
As of this writing, we plan to add 300~kg of PPO to the $1\times10^{6}$~liters of MiniBooNE mineral
oil (300~mg/l).  A preliminary price quote for PPO from the supplier to NOvA is \$250/kg  or \$75k for
scintillator.  The solubility of PPO will allow us to add the entire 300~kg to the MiniBooNE 
10~kl overflow and then introduce that into the main volume by recirculation.  However, it would be
prudent to do this addition in at least two steps by taking the concentration to about 50\% of the 
desired amount and
monitoring detector response with cosmic muons and muon-decay electrons.  We may do recirculation
without the addition of scintillator as a first step, 
as the MiniBooNE oil has not been recirculated since commissioning in 2002.

\subsection{Detector changes}
Our base plan for running with scintillator is only to add scintillator with no other changes.  New
readout electronics could be considered, but are not required for the physics goals set here.  The 
current rate of PMT and failures extrapolated for a 3-year run is not a problem.  We estimate that
the rate of electronics failures over that time period will be covered with our current supply 
of spares.  There will likely be some changes to the computing infrastructure to keep up with
hardware failures and security concerns, but an ``as-needed'' approach is our current plan. 

\subsection{Run plan}
The neutron fraction measurement for oscillation candidates is
statistics limited and $6.5\times10^{20}$ POT is required for our current desired accuracy.  When the MicroBooNE
experiment is running, our assumption is that $2\times10^{20}$~POT/year will be delivered to the Booster Neutrino 
Beamline (BNB).  This sets a 3-year duration for the proposed scintillator phase of MiniBooNE. 
Preparation for running only requires the addition of scintillator (along with modest detector
maintenance), which we estimate will require about 3 months with no beam requirement.  

\section{Conclusions}
The addition of $300$~mg/l of scintillator to the existing MiniBooNE 
mineral oil will allow for the detection and reconstruction of 2.2~MeV $\gamma$  
from neutron-capture.  CC oscillation signal events should have an associated
neutron in less than 10\% of events in contrast to NC background events in which
$\approx 50\%$ have neutrons.   The neutron-capture 
rate for both of these event types can be separately measured in MiniBooNE, 
thus eliminating dependence on neutron production model calculations.  
Therefore, a measurement of neutron-capture in
oscillation events measures the NC backgrounds.  

A measurement of the neutron-fraction in a new appearance oscillation 
search with MiniBooNE will increase the significance of the oscillation
excess, if it maintains in the new data set, to $\approx 5\sigma$.  In practice,
the original oscillation search will be conducted again after the introduction
of scintillator.  With  $6.5\times10^{20}$ POT, the results of this search (before 
neutron capture cuts) should have similar sensitivity as existing search
but with different systematic errors.  Combining this with the neutron
capture analysis will raise the sensitivity to the $5\sigma$ level, perhaps
better, depending on final systematics.

This new phase of MiniBooNE would enable additional important 
studies such as the spin structure of nucleon ($\Delta s$) via NC elastic
scattering, a low-energy measurement of the neutrino flux via the 
$\numu~^{12}C \rightarrow \mu^{-}~^{12}N_\textrm{g.s.}$ reaction, and a test of the
quasielastic assumption in neutrino energy reconstruction.  
This effort will provide traning for  Ph.D.~students and postdocs and 
will yield important, highly-cited results over the next 5 years for 
a modest cost.


\begin{thebibliography}{99}

\bibitem{AguilarArevalo:2007it} 
  A.~A.~Aguilar-Arevalo {\it et al.}  [MiniBooNE Collaboration],
  Phys.\ Rev.\ Lett.\  {\bf 98}, 231801 (2007)
  [arXiv:0704.1500 [hep-ex]].

\bibitem{AguilarArevalo:2008rc} 
  A.~A.~Aguilar-Arevalo {\it et al.}  [MiniBooNE Collaboration],
  Phys.\ Rev.\ Lett.\  {\bf 102}, 101802 (2009)
  [arXiv:0812.2243 [hep-ex]].

\bibitem{AguilarArevalo:2009xn} 
  A.~A.~Aguilar-Arevalo {\it et al.}  [MiniBooNE Collaboration],
  Phys.\ Rev.\ Lett.\  {\bf 103}, 111801 (2009)
  [arXiv:0904.1958 [hep-ex]].

\bibitem{AguilarArevalo:2010wv} 
  A.~A.~Aguilar-Arevalo {\it et al.}  [MiniBooNE Collaboration],
  Phys.\ Rev.\ Lett.\  {\bf 105}, 181801 (2010)
  [arXiv:1007.1150 [hep-ex]].

\bibitem{AguilarArevalo:2012va} 
  A.~A.~Aguilar-Arevalo {\it et al.}  [MiniBooNE Collaboration],
  arXiv:1207.4809 [hep-ex].

\bibitem{Aguilar:2001ty} 
  A.~Aguilar-Arevalo {\it et al.}  [LSND Collaboration],
  Phys.\ Rev.\ D {\bf 64}, 112007 (2001)
  [hep-ex/0104049].


\bibitem{Garvey:1993sg} 
  G.~Garvey, E.~Kolbe, K.~Langanke and S.~Krewald,
  Phys.\ Rev.\ C {\bf 48}, 1919 (1993).

\bibitem{Bass:2009dr} 
  S.~D.~Bass,
  Mod.\ Phys.\ Lett.\ A {\bf 24}, 1087 (2009)
  [arXiv:0905.4619 [hep-ph]].

\bibitem{AguilarArevalo:2010cx} 
  A.~A.~Aguilar-Arevalo {\it et al.}  [MiniBooNE Collaboration],
  Phys.\ Rev.\ D {\bf 82}, 092005 (2010)
  [arXiv:1007.4730 [hep-ex]].

\bibitem{Dharmapalan:2011sa} 
  R.~Dharmapalan [MiniBooNE Collaboration],
  AIP Conf.\ Proc.\  {\bf 1405}, 89 (2011)
  [arXiv:1110.6574 [hep-ex]].

\bibitem{Bugel:2004yk} 
  L.~Bugel {\it et al.}  [FINeSSE Collaboration],
  hep-ex/0402007.

\bibitem{AguilarArevalo:2010zc} 
  A.~A.~Aguilar-Arevalo {\it et al.}  [MiniBooNE Collaboration],
  Phys.\ Rev.\ D {\bf 81}, 092005 (2010)
  [arXiv:1002.2680 [hep-ex]].

\bibitem{Athanassopoulos:1997rn} 
  C.~Athanassopoulos {\it et al.}  [LSND Collaboration],
  Phys.\ Rev.\ C {\bf 56}, 2806 (1997)
  [nucl-ex/9705002].

\bibitem{Athanassopoulos:1997rm} 
  C.~Athanassopoulos {\it et al.}  [LSND Collaboration],
  Phys.\ Rev.\ C {\bf 55}, 2078 (1997)
  [nucl-ex/9705001].

\bibitem{Bodmann:1992ur} 
  B.~Bodmann {\it et al.}  [KARMEN Collaboration],
  Phys.\ Lett.\ B {\bf 280}, 198 (1992).

\bibitem{Kolbe:1999au} 
  E.~Kolbe, K.~Langanke and P.~Vogel,
  Nucl.\ Phys.\ A {\bf 652}, 91 (1999)
  [nucl-th/9903022].

\bibitem{AguilarArevalo:2010bm} 
  A.~A.~Aguilar-Arevalo {\it et al.}  [MiniBooNE Collaboration],
  Phys.\ Rev.\ D {\bf 83}, 052007 (2011)
  [arXiv:1011.3572 [hep-ex]].

\bibitem{AguilarArevalo:2010xt} 
  A.~A.~Aguilar-Arevalo {\it et al.}  [MiniBooNE Collaboration],
  Phys.\ Rev.\ D {\bf 83}, 052009 (2011)
  [arXiv:1010.3264 [hep-ex]].

\bibitem{AguilarArevalo:2009ww} 
  A.~A.~Aguilar-Arevalo {\it et al.}  [MiniBooNE Collaboration],
  Phys.\ Rev.\ D {\bf 81}, 013005 (2010)
  [arXiv:0911.2063 [hep-ex]].

\bibitem{Caballero:2005sj} 
  J.~A.~Caballero, J.~E.~Amaro, M.~B.~Barbaro, T.~W.~Donnelly, C.~Maieron and J.~M.~Udias,
  Phys.\ Rev.\ Lett.\  {\bf 95}, 252502 (2005)
  [nucl-th/0504040].

\bibitem{Martini:2009uj} 
  M.~Martini, M.~Ericson, G.~Chanfray and J.~Marteau,
  Phys.\ Rev.\ C {\bf 80}, 065501 (2009)
  [arXiv:0910.2622 [nucl-th]].

\bibitem{Martini:2010ex} 
  M.~Martini, M.~Ericson, G.~Chanfray and J.~Marteau,
  Phys.\ Rev.\ C {\bf 81}, 045502 (2010)
  [arXiv:1002.4538 [hep-ph]].

\bibitem{Martini:2012fa} 
  M.~Martini, M.~Ericson and G.~Chanfray,
  Phys.\ Rev.\ D {\bf 85}, 093012 (2012)
  [arXiv:1202.4745 [hep-ph]].

\end{thebibliography}
\end{document}